\documentclass[a4paper, amsfonts, amssymb, amsmath, reprint, showkeys, nofootinbib, twoside, superscriptaddress, aps]{revtex4-1}
\usepackage{amsthm}
\usepackage[english]{babel}
\usepackage[utf8]{inputenc}
\usepackage[colorinlistoftodos, color=green!40, prependcaption]{todonotes}
\usepackage{mathtools}
\usepackage{physics}
\usepackage{xcolor}
\usepackage{graphicx}
\usepackage[left=23mm,right=13mm,top=35mm,columnsep=15pt]{geometry} 
\usepackage{adjustbox}
\usepackage{placeins}
\usepackage[T1]{fontenc}
\usepackage{lipsum}
\usepackage{csquotes}
\usepackage{tikz}
\usepackage{bm}
\usepackage{braket}
\usepackage{multirow}
\usepackage{qcircuit}
\usepackage{siunitx}

\usepackage{hyperref}
\hypersetup{
    colorlinks=true,
    linkcolor=blue,
    urlcolor=blue,
    citecolor=blue
    }
\usepackage{cleveref} 

\newcommand{\fidP}[1]{F\left(#1\right)}
\newcommand{\fiddecoh}[1]{F_{\rm decoh.}\left(#1\right)}

\newcommand{\C}{\mathcal{C}}
\newcommand{\D}{\mathcal{D}}
\newcommand{\E}{\mathcal{E}}
\newcommand{\F}{\mathcal{F}}
\newcommand{\h}{\mathcal{H}}
\newcommand{\U}{\mathcal{U}}
\newcommand{\W}{\mathcal{W}}

\usepackage{ulem}

\begin{document}

\title{Extending the Computational Reach of a Superconducting Qutrit Processor}

\author{Noah Goss}
    \thanks{These authors contributed equally to this work. Correspondence should be addressed to \href{mailto:noahgoss@berkeley.edu}{noahgoss@berkeley.edu}.}
    \affiliation{Quantum Nanoelectronics Laboratory, Department of Physics, University of California at Berkeley, Berkeley, CA 94720, USA}
    \affiliation{Applied Mathematics and Computational Research Division, Lawrence Berkeley National Lab, Berkeley, CA 94720, USA}
\author{Samuele Ferracin}
    \thanks{These authors contributed equally to this work. Correspondence should be addressed to \href{mailto:noahgoss@berkeley.edu}{noahgoss@berkeley.edu}.}
    \affiliation{Keysight Technologies Canada, Kanata, ON K2K 2W5, Canada}
\author{Akel Hashim}
    \affiliation{Quantum Nanoelectronics Laboratory, Department of Physics, University of California at Berkeley, Berkeley, CA 94720, USA}
\author{Arnaud Carignan-Dugas}
    \affiliation{Keysight Technologies Canada, Kanata, ON K2K 2W5, Canada}
\author{John Mark Kreikebaum}
    \altaffiliation{Current address: Google Quantum AI, Mountain View, CA 94043, USA}
    \affiliation{Applied Mathematics and Computational Research Division, Lawrence Berkeley National Lab, Berkeley, CA 94720, USA}
    \affiliation{Materials Sciences Division, Lawrence Berkeley National Lab, Berkeley, CA 94720, USA}
\author{Ravi K. Naik}
    \affiliation{Applied Mathematics and Computational Research Division, Lawrence Berkeley National Lab, Berkeley, CA 94720, USA}
\author{David I. Santiago}
    \affiliation{Quantum Nanoelectronics Laboratory, Department of Physics, University of California at Berkeley, Berkeley, CA 94720, USA}
    \affiliation{Applied Mathematics and Computational Research Division, Lawrence Berkeley National Lab, Berkeley, CA 94720, USA}
\author{Irfan Siddiqi}
    \affiliation{Quantum Nanoelectronics Laboratory, Department of Physics, University of California at Berkeley, Berkeley, CA 94720, USA}
    \affiliation{Applied Mathematics and Computational Research Division, Lawrence Berkeley National Lab, Berkeley, CA 94720, USA}
    \affiliation{Materials Sciences Division, Lawrence Berkeley National Lab, Berkeley, CA 94720, USA}

\date{\today}

\begin{abstract}
    Quantum computing with qudits is an emerging approach that exploits a larger, more-connected computational space, providing advantages for many applications, including quantum simulation and quantum error correction. Nonetheless, qudits are typically afflicted by more complex errors and suffer greater noise sensitivity which renders their scaling difficult. In this work, we introduce techniques to tailor and mitigate arbitrary Markovian noise in qudit circuits. We experimentally demonstrate these methods on a superconducting transmon qutrit processor, and benchmark their effectiveness for multipartite qutrit entanglement and random circuit sampling, obtaining up to 3x improvement in our results. To the best of our knowledge, this constitutes the first ever error mitigation experiment performed on qutrits. Our work shows that despite the intrinsic complexity of manipulating higher-dimensional quantum systems, noise tailoring and error mitigation can significantly extend the computational reach of today's qudit processors.
\end{abstract}


\maketitle

\section{Introduction}\label{sec:intro}

\begin{figure*}
    \centering
    \includegraphics[width = \textwidth]{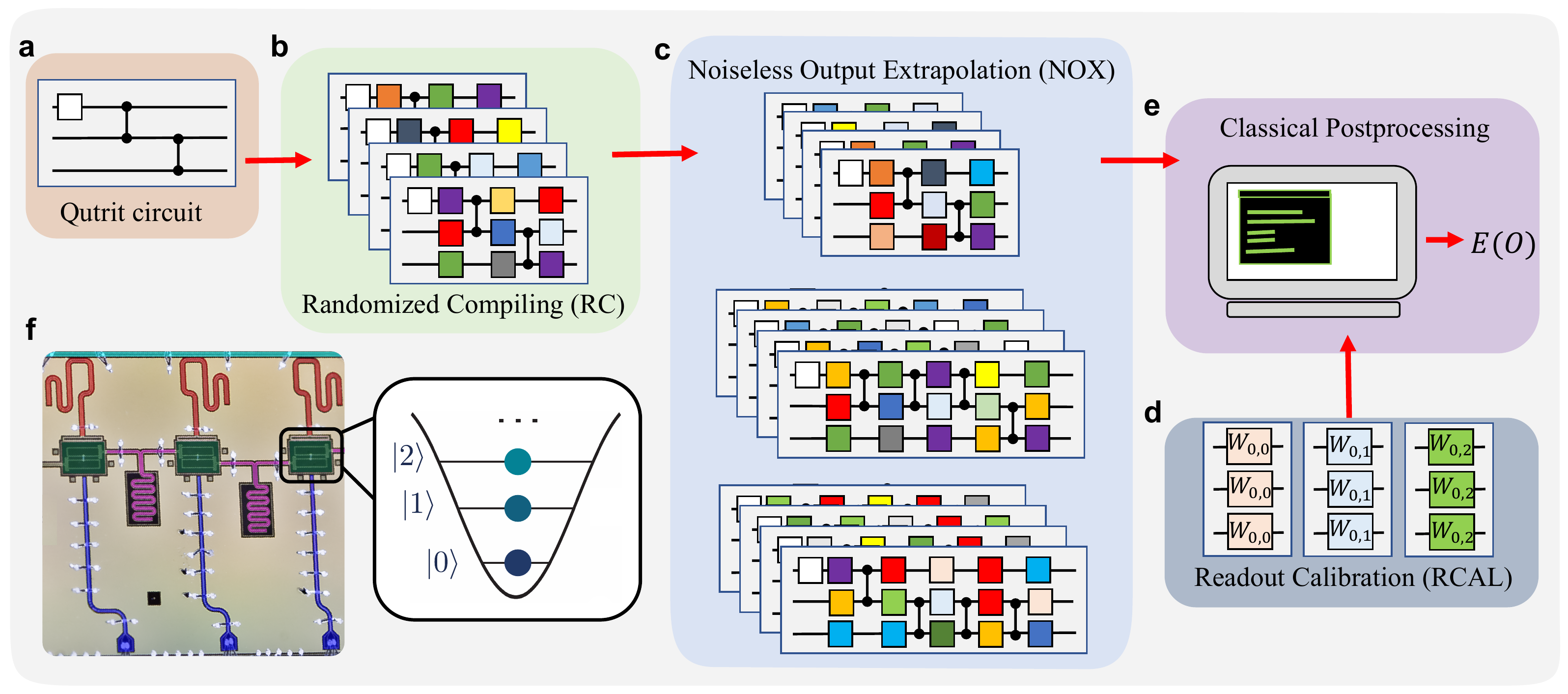}
    \caption{Schematic of error mitigation in qutrit circuits. (a) An arbitrary qutrit circuit is (b) randomly compiled into many logically equivalent copies, effectively twirling the noise of the multi-qutrit gate cycles into stochastic channels. (c) The noise present in the multi-qutrit gates in the circuit is amplified by insertions of the identity. (d) The readout assignment errors are efficiently characterized by measuring the single qutrit confusion matrices. (e) Via classical post-processing, noiseless expectation values are achieved from the combination of RC, NOX, and RCAL. (f) A false colored micrograph of the system of three capacitevly coupled, fixed-frequency transmon qutrits used in the experiment. }
    \label{fig:fig1}
\end{figure*}

In the Noisy Intermediate-Scale Quantum (NISQ) era~\cite{preskill2018quantum}, the ability to effectively couple many quantum two-level systems (qubits) together has led to experimental demonstrations of certain tasks that challenge the limits of current classical capabilities \cite{arute2019quantum,zhong2020quantum,PhysRevLett.127.180501, photonic-advantage}. The computational power of these near-term devices can potentially be further boosted by leveraging the innate multi-level structure to encode quantum information in the larger and more connected Hilbert space of $d$-level systems (qudits) \cite{quditalgs1, PhysRevA.96.012306, nikolaeva2022efficient, PhysRevLett.94.230502, simplifyinglogic,singlequdit}. Coherent control of higher level quantum systems has been demonstrated in superconducting circuits \cite{scrambling, PhysRevLett.126.210504, goss, PhysRevLett.130.030603,roy2022realization, cao2023emulating, Litteken:2023qzj}, trapped ions \cite{ringbauer2021universal, ringbauer2022}, and in photonic circuits \cite{lanyon2008manipulating, photonic-qudit}. 

 The simplest and most immediately experimentally viable member of the qudit family is the the quantum three-level system, or qutrit. Qutrits can yield specific advantages in quantum simulations, where they are a natural platform for studying spin-1  physics \cite{wang2023dissipative, PhysRevX.5.021026} and robust and resource efficient for simulating high-energy phenomena \cite{sqed-simulation, scrambling}. Additional applications include improvements in quantum cryptography \cite{PhysRevLett.85.3313, PhysRevLett.88.127901}, communication \cite{PhysRevLett.89.240401}, compactly synthesizing multi-qubit gates \cite{10.1145/3307650.3322253, federovtoffoli,quantumand, nguyen2022programmable, galda2021implementing, hill2021realization}, and for improving qubit readout \cite{PhysRevX.10.011001,esp, chen2022transmon}. Eventually, qutrits are expected to provide significant advantages for quantum error correction via improved error thresholds \cite{PhysRevA.87.062338, qudit-toric-codes-threshold, PhysRevA.91.042331, ma2023nonpauli}, errors tailored to erasure \cite{muralidharan_zou_li_wen_jiang_2017, kubica2022erasure}, enhanced fault tolerance \cite{PhysRevX.2.041021,PhysRevLett.113.230501}, and compact encodings of both logical qutrits \cite{PhysRevA.97.052302} and logical qubits \cite{PhysRevLett.116.150501, li2023autonomous}.

 While fault tolerance remains the ultimate goal, alternative efforts in qubit devices to mitigate and extrapolate expectation values beyond the noise present in the system have garnered interest lately due to the lack of an increase in hardware requirements and overall feasibility \cite{LB17,TBG17,EBL18,KandalaEtal19,SQCBY20,GTHLMZ20,LRMKSZ20,ECBY21,KWYMGTK21,Koczor21,HNdYB20,HMOLRBWBM21,BHdJNP21,SongEtAl19,ZhangEtAL20,CACC20}. Notably, recent works have demonstrated that error mitigation protocols can prove effective for large scale qubit experiments, allowing an exciting pathway to high-fidelity results in the near term for problems of interest \cite{EBL18,KandalaEtal19,SongEtAl19,ZhangEtAL20,SQCBY20,CACC20,HNdYB20,BHdJNP21,KWYMGTK21,Ewout22,ferracin2022efficiently}. Recent investigations have also explored the compatibility of noise tailoring, error mitigation, and quantum error correction \cite{jain2023improved,hashim2021randomized}. However, there is a dearth of similar studies for qudit devices, which are now approaching the maturity to scale to larger experimental sizes, but are afflicted by more complicated noise processes \cite{noise-sensor}. This opens the door to an interesting question:  can the computational power of contemporary devices be augmented by enlarging their Hilbert space as qudits, while at the same time retaining the ability to generate noiseless expectation values in this more complex noise environment?

In this work, we attempt to answer this question, and show that error mitigation can be utilized to significantly extend the existing resources of a superconducting qutrit processor. More explicitly, we introduce two techniques for tailoring and mitigating noise in qutrit circuits: Randomized Compiling (RC) \cite{wallman2016noise,hashim2021randomized} and Noiseless Output Extrapolation (NOX) \cite{ferracin2022efficiently}. We then explore the performance of these techniques in a variety of multi-qutrit experiments using fixed frequency transmons. In particular, we study state tomography of a three-qutrit Greenberger-Horne-Zeilinger (GHZ) state, as well as random circuit sampling (RCS) with two and three qutrits. We find that in all cases, despite the more complex noise environment, our results benefit greatly from these protocols, achieving up to a 3 times improvement in fidelity. Our work is the first to experimentally demonstrate that error-mitigation can be effectively implemented on qutrit platforms, and it paves the way to scaling near-term, large-scale qutrit computations.

\section{Computing, Twirling, and Mitigating with Qutrits}\label{sec:qutrit}
\begin{figure*}
    \centering
    \includegraphics[width = \textwidth]{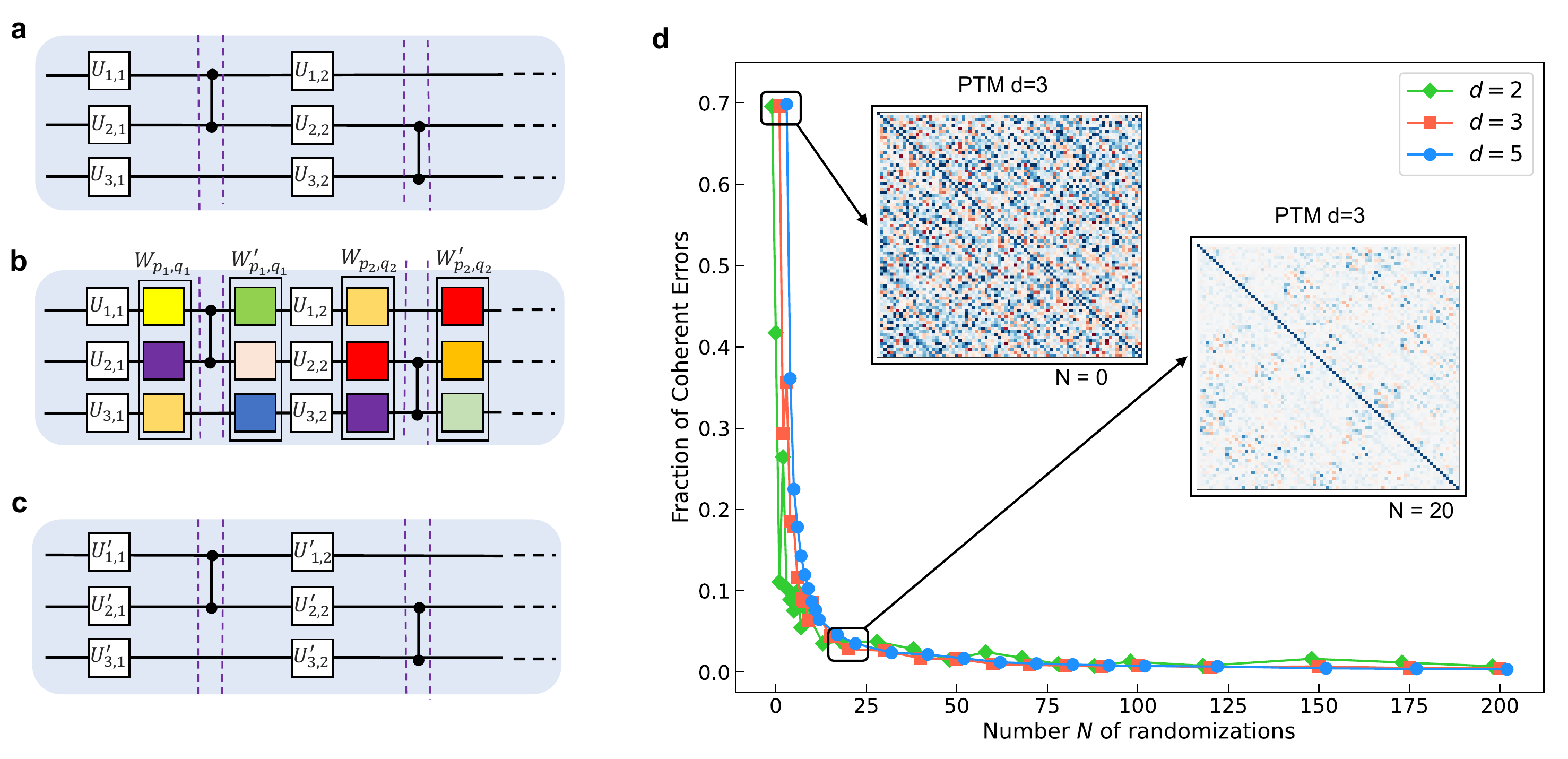}
    \caption{Randomized Compiling (RC) for qudit circuits. (a) The input circuit, which alternates cycles of one- and multi-qudit gates. (b) Random Weyl gates ($W_{\bar{p}_j,\bar{q}_j}$) and their inverses ($W_{\bar{p}_j,\bar{q}_j}'=W_{\bar{p}_j,\bar{q}_j}^\dagger H_jW_{\bar{p}_j,\bar{q}_j}$) are added before and after every cycle of multi-qudit gates $H_j$. (c) Before executing the circuit, the extra gates are recompiled into the native one-qudit gates. In this way, the returned circuit has the same depth as the input one. (d) Numerical study of the fraction of coherent errors under RC randomizations in randomly generated two-qudit Pauli Transfer Matrices (PTM) in $d\in\{2,\:3,\:5\}$. The formula used to quantify the fraction of coherent errors is derived in Appendix~\ref{sec:coh_fraction}. All PTMs are chosen to initially have 70\% of their errors be coherent. As it can be seen, the coherent errors decrease exponentially with $N$, confirming the validity of Eq.~\ref{eq:hoeffdings}. Furthermore, its value does not depend on $d$. This demonstrates that for an equal suppression of coherent errors, RC does not require additional twirls in higher dimensions. The inset PTMs visualize the suppression of off-diagonal terms as a function of $N$.}
    \label{fig:fig2}
\end{figure*}

\textit{Computing---}With qutrits, information is encoded in the three energy levels that correspond to the computational basis states $\ket{0}$, $\ket{1}$, and $\ket{2}$. As for qubits, gates are unitary operators relative to the computational basis. An important set of qutrit gates is the Weyl operators, which are a generalization of the of the qubit Pauli group in a higher dimensional Hilbert space. The action of the Weyl gates is specified by the operators $W_{p,q}=\omega^{-pq/2}Z^pX^q$, where $X$ and $Z$ are defined by their action on the basis state $\ket{n}$ as $X\ket{n}=\ket{n\oplus1}$ and $Z\ket{n}=\omega^{n}\ket{n}$ and $\omega=e^{2i\pi/3}$. Like the Pauli gates for qubit systems, the Weyl operators form a unitary 1-design \cite{GSNW22} and are normalized by the qutrit Clifford group.

Natural generalizations of two-qubit gates can also be constructed. For example, the two-qutrit analogue of the controlled-Z (CZ) gate is defined as,
\begin{equation}
\label{eq:cz}
    \textnormal{CZ} = \sum_{n=0}^{n=2}\ketbra{n}{n}\otimes Z^n .
\end{equation}
Both CZ and its inverse CZ$^\dag$ are universal, Clifford entangling gates that can be performed on our system \cite{goss}. Critically, the natural adoption of twirling and mitigation methods that have been developed for qubits is only possible through the ability to implement two-qutrit Clifford gates.

\textit{Twirling---}Being a unitary 1-design, Weyl operators can be used to \textit{twirl} noise \cite{GSNW22,hashim2021randomized,FKD18,FMMD21}---that is, to transform arbitrary Markovian noise processes into stochastic channels of the form
\begin{equation}
    \label{eq:twirled-noise}\W(\rho)=\sum_{\bar{p},\bar{q}}^{4^n}\textrm{prob}(W_{\bar{p},\bar{q}})W_{\bar{p},\bar{q}}\:\rho \:W_{\bar{p},\bar{q}}^\dagger\:,
\end{equation}
where $\rho$ is an $n$-qutrit state, $W_{\bar{p},\bar{q}}=\otimes_{i=1}^nW_{q_{k_i},p_{k_i}}$ is a tensor product of one-qutrit Weyl operators, and $\textrm{prob}(W_{\bar{p},\bar{q}})$ is the probability that a Weyl error $W_{\bar{p},\bar{q}}$ occurs. Twirling noise processes can significantly improve the performance of noisy devices \cite{hashim2021randomized}. Indeed, while coherent errors can accumulate quadratically in the number of noisy gates, stochastic channel accumulate linearly and dramatically lower worst-case error rates~\cite{hashim2022benchmarking}.

In our experiments, we twirl the noise using the RC protocol from Ref.~\cite{wallman2016noise} generalized to qutrits. To illustrate this procedure, let us consider a ``target'' circuit of the type in Fig.~\ref{fig:fig2}\textcolor{blue}{a}, i.e., a circuit that alternates between cycles of single-qutrit gates (represented by Completely Positive Trace-Preserving, or CPTP, maps $\U_j$) and cycles of multi-qutrit gates (represented by CPTP maps $\h_j$), implementing the operation
\begin{equation}
    \C=\U_{m+1}\h_m\U_m\cdots\h_1\U_1\:.
\end{equation}
Expressing a noisy implementation of $\U_j$ (respectively $\h_j$) as $\D_j\U_j$ (respectively $\F_j\h_j$), a noisy implementation of this circuit performs the operation
\begin{equation}
    \widetilde{\C}=\U_{m+1}\E_m\h_m\U_m\cdots\E_1\h_1\U_1\:,
\end{equation}
where $\E_j=\D_{j+1}\h_j\F_j\h_j^{-1}$ is the combined noise of $\U_{j+1}$ and $\h_j$. Analogous to the original protocol, to perform RC we recompile randomly-chosen Weyl gates and their inverses into the cycles of one-qubit gates (Fig.s~\ref{fig:fig2}\textcolor{blue}{b}, \ref{fig:fig2}\textcolor{blue}{c}). Implementing the circuit $N>1$ times with different choices of random Weyl gates and averaging over the various implementations is equivalent to implementing a circuit
\begin{equation}
    \widetilde{\C}_\textrm{RC}=\U_{m+1}\W_m\h_m\U_m\cdots\W_1\h_1\U_1
\end{equation}
afflicted by stochastic noise processes $\W_j$ of the type in Eq.~\ref{eq:twirled-noise}, up to statistical fluctuations that decrease exponentially with $N$. Specifically, denoting by $\widetilde{\C}_1,\ldots, \widetilde{\C}_N$ the $N$ circuits with random Weyl operators, Hoeffding's inequality \cite{H63} ensures that for every state $\rho$, observable $O$, and positive number $\epsilon<1$, we have
\begin{align}
    \label{eq:hoeffdings}&\textrm{prob}\bigg(\big| E(O|\widetilde{\C}_\textrm{RC})-\frac{1}{N}\sum_{k=1}^NE(O|\widetilde{\C}_k)\big|<\epsilon\big)\nonumber\\
    &\geq1-\textrm{exp}\big(-2\epsilon^2N\big)\:,
\end{align}
where $E(O|\C)=\textrm{Tr}\big[O\C\big(\rho\big)\big]$ is the expectation value of $O$ at the end of a circuit $\C$. 
Importantly, the r.h.s.~of the above inequality does not depend on the dimension of the Hilbert space. Hence, achieving the desired approximation level requires implementing the same number of circuits for qutrits as for qubits (Fig.~\ref{fig:fig2}\textcolor{blue}{d}).\\

\textit{Mitigating---}While implementing RC alone leads to significant performance gains, employing it in tandem with other error-mitigation protocols leads to even larger gains. In our experiments, we utilize RC in combination with NOX, a protocol designed to mitigate stochastic errors afflicting cycles of gates. In its simplest version, NOX requires implementing the target circuit $\widetilde{\C}_\textrm{RC}$ alongside several copies of it. Each copy implements the same computation as $\widetilde{\C}_\textrm{RC}$, but the noise of one of its cycles is amplified in a controlled way. Specifically, the $j$-th copy implements the operation
\begin{align}
    &\widetilde{\C}^{(j)}_{\textrm{RC}}:=\\
    &\U_{m+1}\W_m\h_m\U_m\cdots \left(\W_j\right)^{\alpha_j} \h_j\U_j\cdots\W_1\h_1\U_1\:,\nonumber
\end{align}
where $\alpha_j>1$ is an integer specified by the user.
By combining the outputs of the target circuit and those of the various copies, NOX returns the quantity
\begin{equation}
    E_\textrm{NOX}(O)=E(O|\widetilde{\C}_\textrm{RC})-\sum_{j=1}^m\frac{E(O|\widetilde{\C}_\textrm{RC})-E(O|\widetilde{\C}^{(j)}_\textrm{RC})}{\alpha_j-1}\:.
\end{equation}
$E_\textrm{NOX}(O)$ is an estimator of the correct result, $E(O|\C)$, and its bias is quadratically smaller than that of $E(O|\widetilde{\C})$ \cite{ferracin2022efficiently}.

In order to perform noise amplification, in our experiments we utilize Unitary Folding~\cite{GTHLMZ20}. In more detail, given a noisy cycle $\W\h$ and a number $\alpha$ such that $\h^{\alpha+1}=\h$, we replace $\W\h$ with $(\W\h)^{\alpha+1}$. In our experiments, the cycle $\h$ is a tensor product of CZ$^\dag$ gates. Hence, the noise process $\W$ is dominated by diagonal errors (i.e., by errors of the type $W_{\bar{p}, \bar{0}}$) and $\W\h\approx\h\W$. 
This leads to $(\W\h)^{\alpha+1}\approx\W^{\alpha+1}\h$, which provides the desired noise amplification. More sophisticated techniques can be used to amplify noise when $\W\h\not\approx\h\W$~\cite{ferracin2022efficiently}, in combination with cycle-based noise reconstruction techniques~\cite{CD22}.

In addition to RC and NOX, we employ Readout Calibration (RCAL) to mitigate measurement noise~\cite{trueq}. RCAL requires implementing three simple circuits (the first with a $W_{0,0}$ gate on every qutrit, the second with a $W_{0,1}$ gate, and the third one with a $W_{0,2}$ gate) to estimate the probabilities of state-dependent readout errors---i.e., the probabilities that an output $s\in\{0,1,2\}$ is incorrectly reported as $s'\neq s$. It then uses this information to efficiently suppress readout errors by inverting confusion matrices.

\section{Experiment}\label{sec:exp}
\begin{figure*}
    \centering
    \includegraphics[width = \textwidth]{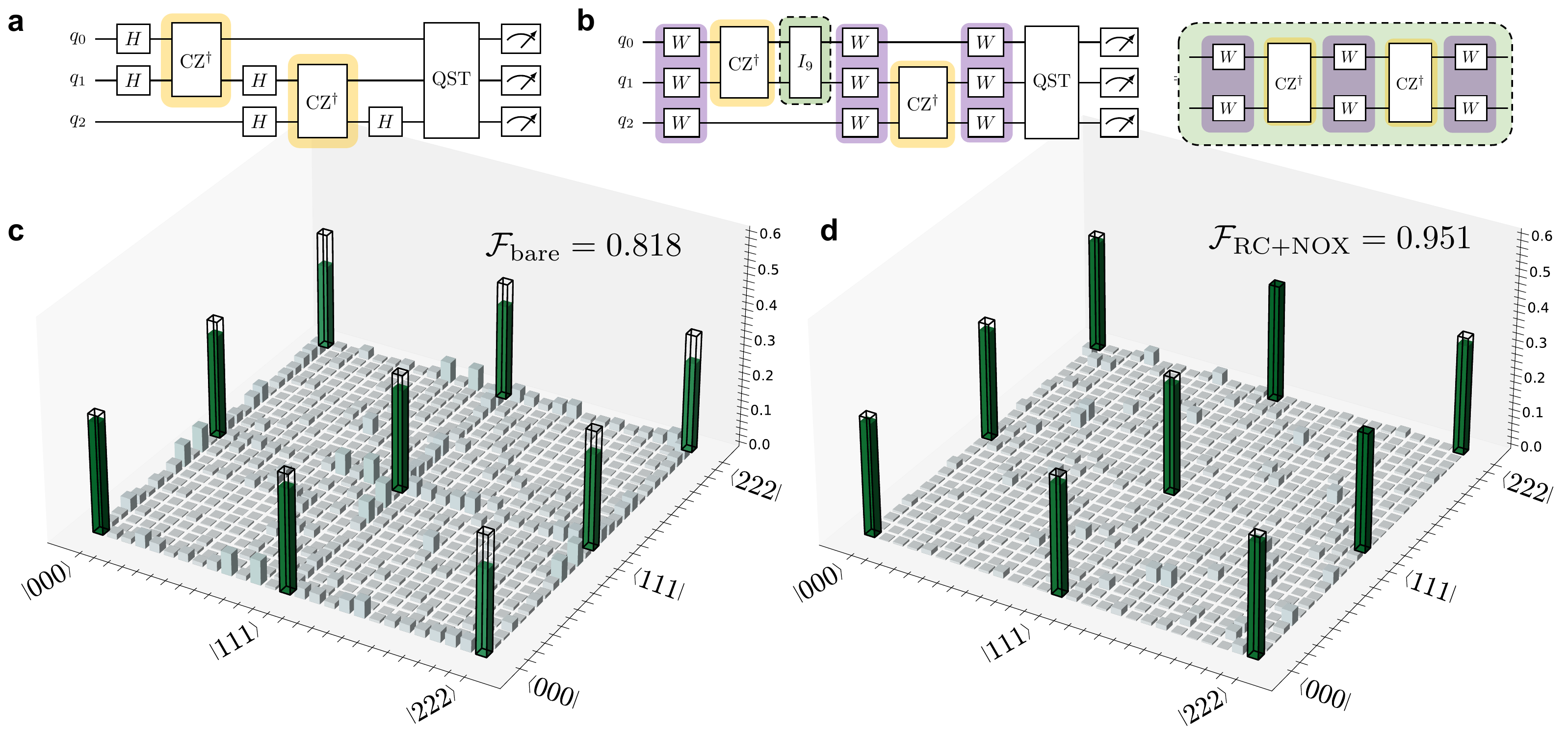}
    \caption{Experimentally reconstructed qutrit GHZ density matrices. (a) The circuit diagram for experimentally reconstrucing the density matrix of a 3-qutrit GHZ state. The state can be efficiently generated with two-qutrit $CZ^\dag$ gates (yellow) and single qutrit Hadamard gates (H). (b) An example circuit diagram with RC+NOX, including Weyl twirling (purple), where the identity insertion (green) is placed  after the CZ$^{\dag}$ (yellow) between $q_0$ and $q_1$. An example of a twirled decomposition of the identity (green) is provided to the right. (c) The experimentally reconstructed density matrix (plotting $|\rho|$) of the 3-qutrit GHZ state with only readout assignment errors corrected. The ideal density matrix is shown with a black outline. (d) The density matrix generated by the noiseless expectation values produced by RC+NOX.}
    \label{fig:exp_ghz}
\end{figure*}
We test our ability to effectively twirl and extrapolate noiseless results in two sets of multi-qutrit experiments. In the first experiment, we use state tomography to reconstruct a three-qutrit GHZ state. In the second experiment, we perform random circuit sampling (RCS) for two and three qutrits at a variety of depths. In both experiments, we find significant improvements in our results via the combination of RC and NOX. 

Our experimental device consists of fixed-frequency transmons, with fixed coupling mediated by coplanar waveguide resonators. Single-qutrit gates are performed via Rabi oscillations and virtual Z gates in two-level subspaces of the qutrit, and two-qutrit gates are performed via a tunable cross-Kerr entangling interaction. More information on how we perform single  and two-qutrit gates can be found in refs.~\cite{scrambling, PhysRevLett.126.210504} and ref.~\cite{goss} respectively and additional device characterization is available in the Appendix~\ref{sec:characterization}.

\subsection{Multipartite Qutrit Entanglement}
Experimental demonstrations of multipartite entanglement have been instrumental in demonstrating that local realism can be violated by quantum mechanics \cite{hensen2015loophole}. So far, studies of multipartite entanglement have mostly focused on coupled qubits, and experimental demonstrations of multipartite qutrit entanglement have only been performed in a few cases \cite{scrambling,PhysRevApplied.17.024062}. In this work, we generate a maximally entangled state on a $D=27$ dimensional Hilbert space using only three transmon qutrits. Specifically, we realize the qutrit GHZ state $\ket{\Psi}_{\textnormal{GHZ}} = \frac{1}{\sqrt{3}}(\ket{000} + \ket{111} + \ket{222})$. 

To characterize this highly-entangled state, we perform tomography on the full three-qutrit Hilbert space \cite{PhysRevLett.105.223601}, requiring a total of 729 circuits. We measure an informationally-complete set of projections to experimentally reconstruct the density matrix for $\rho = \ketbra{\Psi}{\Psi}_{\textnormal{GHZ}}$, finding a state fidelity of $\mathcal{F}= \textnormal{Tr}(\sqrt{\sigma}\rho \sqrt{\sigma}) = 0.818$, where $\sigma$ is the ideal density matrix. Even in the unmitigated case, our work represents, to the best of our knowledge, the highest fidelity exploration of multipartite qutrit entanglement to date.

Next, we perform the same experiment using RC and NOX, requiring a total of 43740 (729$\times$20$\times$3) circuits, where we use 20 logically-equivalent compilations of the bare circuit and for each of the identity insertion for the two different two-qutrit gate cycles. We find a mitigated state fidelity of $\mathcal{F}= 0.951$, resulting in a greater than 3x reduction in infidelity compared to the unmitigated case. The circuits and the experimentally-reconstructed density matrices can be found in Fig.~\ref{fig:exp_ghz}. Finally, we purify the reconstructed state measured with RC (no NOX) \cite{ville2021leveraging} by numerically finding the nearest density matrix that is idempotent ($\rho^2 = \rho$) \cite{purification}, which improved the state fidelity from 0.912 to 0.998 (see Appendix~\ref{section:purification}). These results demonstrate the power of RC to tailor coherent errors to purely stochastic channels.

\subsection{Qutrit Random Circuit Sampling}
\begin{figure}
    \centering
    \includegraphics[width = 0.49\textwidth]{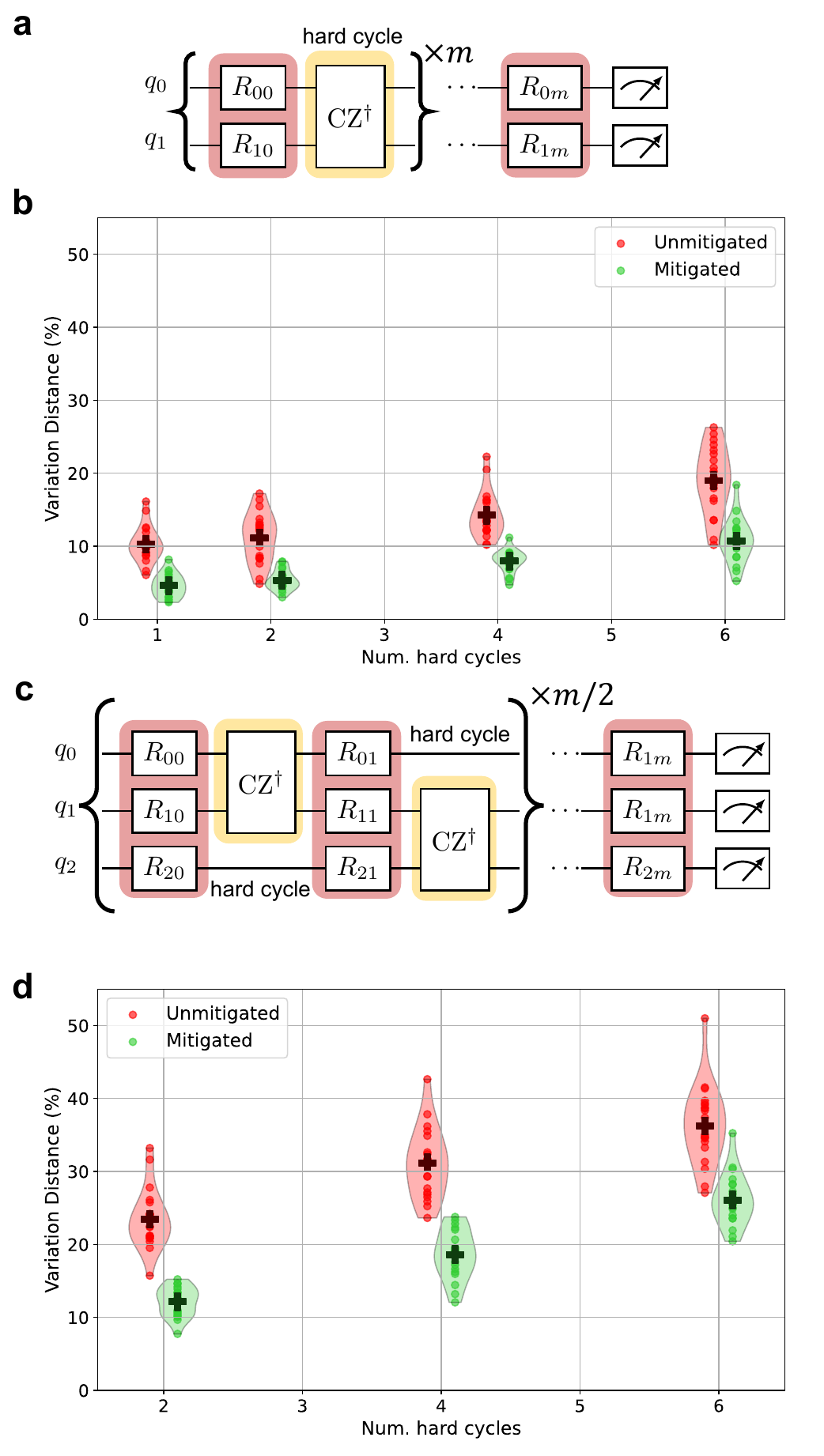}
    \caption{Random circuit sampling with qutrits. (a) The circuits for two-qutrit random circuit sampling. CZ$^\dag$ (yellow) hard cycles are interleaved with Haar random SU(3) gates, $R_{ij}$ (red). (b) Violin plots showing the distribution of the variation distances for the 20 RCS instances with (green) and without (red) RC+NOX at depths $m\in \{1,2,3,6\}$, with mean values marked by crosses. (c) Circuits for three-qutrit RCS. (d) Results for three-qutrit RCS at $m \in \{2,4,6 \}$.} 
    \label{fig:exp_RCS}
\end{figure}

RCS with qubits has recently garnered significant interest due to its role in demonstrating quantum advantage \cite{arute2019quantum,PhysRevLett.127.180501}, and it has been speculated that RCS may also have pragmatic use cases \cite{aaronson2023certified, gokhale2022supercheq}. In this work, we extend the task of RCS to qutrits, and study the two and three-qutrit experiments at a number of depths. Notably, leveraging qutrits for RCS in principle could allow one to probe the regime of quantum supremacy with significantly fewer qutrits ($\sim 35$) than qubits ($\sim 55$).

In both the two and three-qutrit RCS experiments, Haar-random single qutrit gates are interleaved with CZ$^\dag$ gates (hard cycles). We study 20 separate RCS instances for each circuit depth. In Fig.~\ref{fig:exp_RCS}, the bare RCS results are compared to the mitigated results measured using RC (20 randomizations) and NOX. To quantify circuit performance, we calculate the variation distance from the ideal trit string distributions, defined as
\begin{equation}
    \textnormal{VD} := \frac{1}{2}\sum_{\Bar{s}}|p_{\textnormal{id}}(\Bar{s}) - p_{\textnormal{exp}}(\Bar{s})|
\end{equation}
where $\Bar{s}\in (0,1,2)^{\otimes n}$ and $p_{\textnormal{id}}$ and $p_{\textnormal{exp}}$ are the ideal and experimentally measured trit string distributions, respectively. The best results for the mitigated case were found with 3 insertions of the identity (see Appendix \ref{sec:id_inser}). Notably, when employing RC+NOX in both the two and three-qutrit RCS experiments, the variation distances at depth 6 were comparable to the unmitigated case at depth 2, and at all depths we found at least a 30\% fractional improvement in our results with RC+NOX.

\section{Conclusions}\label{sec:conclusions}
We introduced generalized versions of two powerful methods for tailoring and mitigating noise in contemporary qudit systems: randomized compiling and noiseless output extrapolation. We tested the efficacy of these methods at generating noiseless expectation values on a system of three coupled transmon qutrits. Specifically, we explored the experimentally reconstructed density matrix of a 3-qutrit GHZ state and the problem of multi-qutrit random circuit sampling. We found that despite the more complex noise environment, added noise sensitivity, and more difficult control requirements, our protocols proved to be a powerful tool for significantly improving our results for all of the aforementioned experiments. Specifically, we effectively tailored coherent errors to stochastic errors on both gate and spectator qutrits via the first demonstration of randomized compiling for qudit dimension $d\geq 3$, and  extrapolated beyond the noise in our system via the first demonstration of noiseless output extrapolation in $d \geq 3$.

As higher-dimensional quantum devices begin to mature and compete with qubit systems, the ability to perform longer-depth algorithms without significant errors will be critical to convincing the community of their feasibility and scalability. To this end, our work opens the door to explore many of the advantages leveraged by qudit devices in both quantum algorithms and gate based quantum simulation in the near term on contemporary devices. 
\section*{Acknowledgements} \label{sec:acknowledgements}
This material is based upon work supported by the National Science Foundation under Grant No.~2210391. Additional support was provided by the Office of Advanced Scientific Computing Research, Testbeds for Science program, Office of Science of the U.S. Department of Energy under Contract No. DE-AC02-05CH11231. A.H.~acknowledges financial support from the Berkeley Initiative for Computational Transformation Fellows Program.

N.G. acknowledges Long Nguyen, Karthik Siva, and Brian Marinelli for useful discussions. S.F acknowledges Joel Wallman and Anthony Chytros for useful discussions.

S.F.~and A.C.D.~have a financial interest in Keysight Technologies and the use of True-Q software~\cite{trueq}. All other authors declare no competing interests.
\newpage
\appendix
\setcounter{table}{0}
\renewcommand{\thetable}{A\arabic{table}}

\setcounter{figure}{0}
\renewcommand{\thefigure}{A\arabic{figure}}
\renewcommand{\theHfigure}{A\arabic{figure}}

\begin{figure*}
    \centering
    \includegraphics[width = \textwidth]{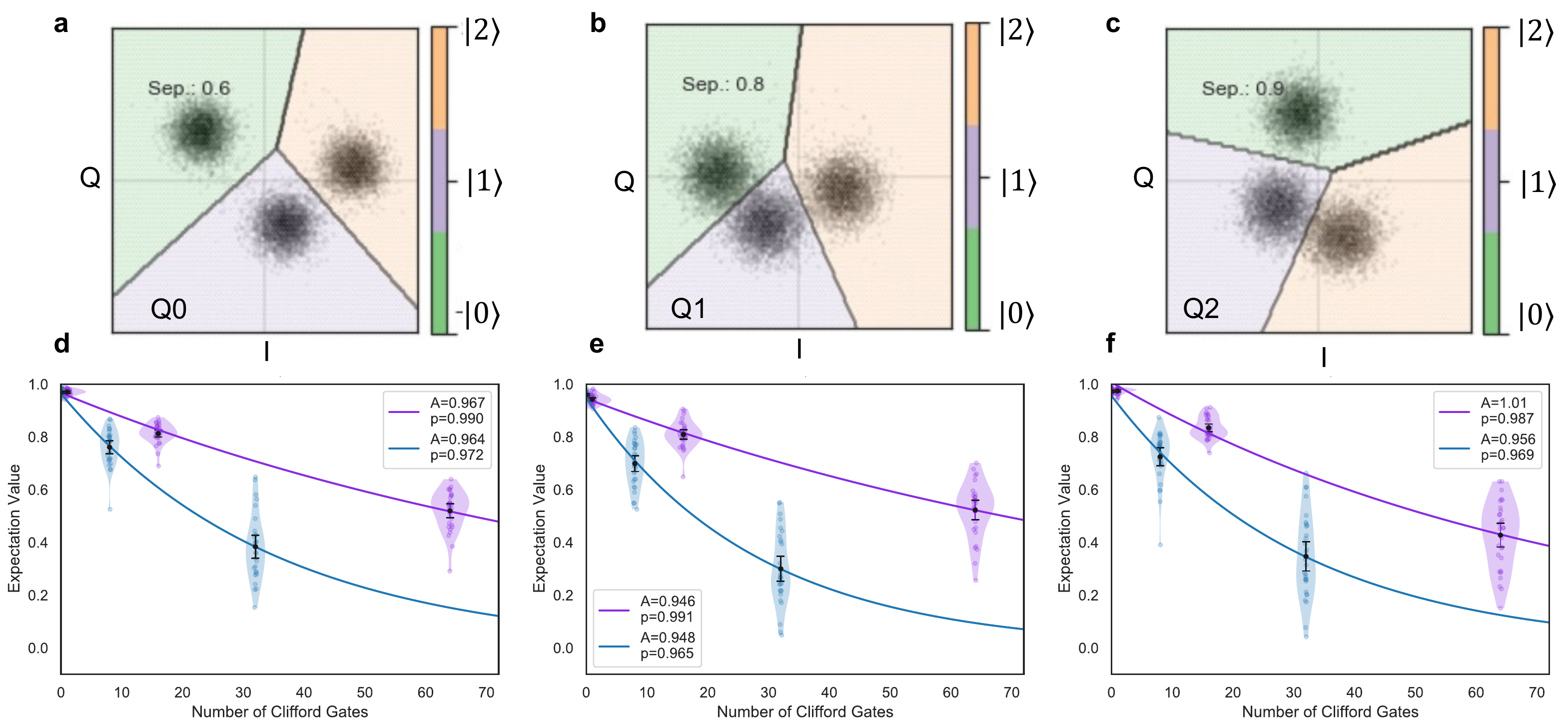}
    \caption{(a)-(c) An example of state discriminated for qutrit dispersive readout on the device with arbitrary units in the IQ plane. (d)-(f) Isolated qutrit randomized benchmarking data (purple) compared to the simultaneous case (blue). $A$ and $p$ are the values extracted from fitting the exponential decay $Ap^{-x}$, where $x$ is the number of qutrit Clifford gates. } 
    \label{fig:S0}
\end{figure*}

\begin{figure*}
    \centering
    \includegraphics[width = 0.9\textwidth]{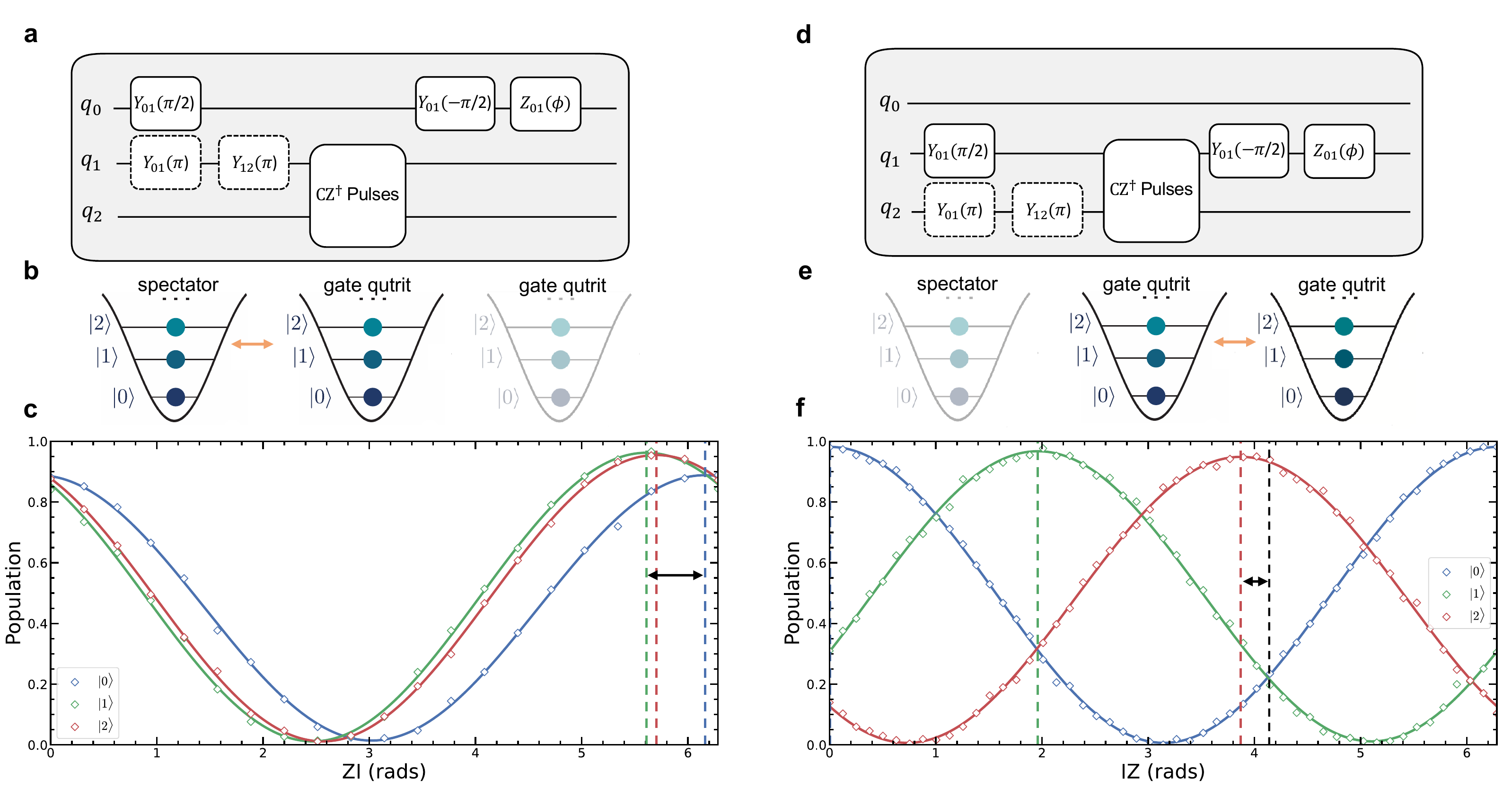}
    \caption{Entangling Phase Characterization. (a) The quantum circuit for characterizing the entangling phase (in a single subspace of the qutrit) between the spectator and gate qutrit given different control states of the gate qutrit. (b) A schematic drawing of the qutrits in the system and their involvment in the experiment. (c) The characterized entangling phase given different control states of the gate qutrit. Note the spread in phase of the sinuosoids demarcated by the black arrow visualizes the coherent error present. (d) The quantum circuit for characterizing the entangling phase (in a single subpace of the qutrit) between the two gate qutrits. (e) A schematic drawing of the qutrits in the system. (f) The characterized entangling phase between the two gate qutrits. Note the uneven spacing in the phase of the sinuosids demarcated by the black arrow visualizes the coherent error present.}
    \label{fig:S1}
\end{figure*}

\section{Device Parameters and Coherence}\label{sec:characterization}
The device employed in these experiments is an 8 qutrit ring with fixed-frequency transmons and fixed-frequency couplings mediated by co-planar waveguide resonators. The experiments are performed on a subset of three transmons on the device and the relevant device parameters and coherences are found in Table~\ref{tab:single_qutrit_parameters}. The results of isolated and simultaneous qutrit randomized benchmarking can also be found in Fig.~\ref{fig:S0}.

\begin{table}[h!]
  \centering
  \resizebox{0.75\columnwidth}{!}{
  \begin{tabular}{|l||r|r|r|r|r|r|r|r|r|r|r|}
    \hline
    {} & Q0  & Q1 & Q2 \\
    \hline
    \hline
    Qubit freq.~(GHz)& 5.299 & 5.362 & 5.523 \\
    Anharm. (MHz)& -272 & -275 & -271 \\
    $T_1^{01}$ ($\mu$s)& 60(13) & 64(20) & 47(14) \\
    $T_1^{12}$ ($\mu$s)& 36 (4) & 33(4) & 34(5) \\
    $T_{2e}^{01}$ ($\mu$s)& 69 (7) & 82(12) & 61(14) \\
    $T_{2e}^{12}$ ($\mu$s)& 32 (5)& 33(4) & 32(6) \\
    $\mathcal{F}_{\textnormal{RO}},P(0|0)$ & 0.994& 0.991& 0.986 \\
    $\mathcal{F}_{\textnormal{RO}},P(1|1)$ & 0.979& 0.953& 0.943 \\
    $\mathcal{F}_{\textnormal{RO}},P(2|2)$ & 0.974& 0.943& 0.951 \\
    $\mathcal{F}_{\textnormal{RB,Iso}}$ & 0.991& 0.992& 0.989 \\
    $\mathcal{F}_{\textnormal{RB,Sim}}$ & 0.975& 0.969& 0.972 \\
    $\alpha_{11}$ (MHz)  & 0.10 & 0.10 & \\
     &  & 0.16 & 0.16\\
    $\alpha_{12}$ (MHz)  & 0.60 & 0.60 &\\
       & &0.41 &0.41\\
    $\alpha_{21}$ (MHz)  & -0.44& -0.44&\\
       & &-0.16 &-0.16\\
    $\alpha_{22}$ (MHz)  & 0.36& 0.36&\\
       & &0.49 &0.49\\
    \hline
  \end{tabular}}
\caption{Measured properties of qutrits for the experimental device employed in this work. Here, the $\alpha_{ij}$ terms are the strength of the static cross-Kerr coupling between the qutrits.}\label{tab:single_qutrit_parameters}
\end{table}

\section{Qutrit Gate Characterization}
The two-qutrit CZ$^\dag$ gates used in this experiment utilize a dynamic cross-Kerr entangling interaction with interleaved echo pulses in the $\{\ket{1},\ket{2} \}$ subspace of the qutrit which shuffle entangling phases. The entangling phases in the two-qutrit Hilbert space are mediated by this cross-Kerr coupling, where we define the cross-Kerr Hamiltonian as:
\begin{align}
    \mathcal{H}_{cK} &= \alpha_{11}\ketbra{11} + \alpha_{12}\ketbra{12} \nonumber \\
                     &+ \alpha_{21}\ketbra{21} + \alpha_{22}\ketbra{22} ,
\end{align}
where in the static case, the strength of the cross-Kerr terms are defined as $\alpha_{ij}/\hbar = (E_{ij} - E_{i0}) - (E_{0j} - E_{00})$. In the case of the driven cross-Kerr coupling, we utilize simultaneous Stark shifts to mediate and tune the cross-Kerr coupling between the two qutrits.  For additional reading on the gate mechanism and tune up, readers are directed to Ref.~\cite{goss}.

\subsection{Coherent Errors on Gate-Involved Qutrits}
In our work, we tune-up two-qutrit unitaries in the form of the qutrit CZ$^\dag$ gate. This unitary takes the following form,
\begin{align}
    U_{CZ^\dag} &= e^{4\pi i /3}( \ketbra{11}{11} + \ketbra{22}{22}) \nonumber \\ 
                &+ e^{2\pi i /3}( \ketbra{12}{12} + \ketbra{21}{21}) \nonumber \\
                &+ \sum_{j=0}^2 (\ketbra{0j}{0j} + \ketbra{j0}{j0}) , 
\end{align}
which requires four precisely calibrated entangled phases. Despite achieving relatively high two-qutrit gate fidelities \cite{goss}, a sizable portion of our error budget can be caused by coherent calibration errors in the two-qutrit gate. As coherent errors can accumulate up to quadratically (versus linearly for stochastic errors), randomized compiling is especially powerful at mitigating the impact of these errors \cite{hashim2021randomized}. To visualize and characterize the coherent error in terms of entangling phase on the gate-involved qutrits, we perform the experiment shown in Figs.~\ref{fig:S1} d-f. 

\subsection{Coherent Errors on Spectator Qutrits}
In the experiments performed in this work, it is necessary to be aware of errors occurring both on the qutrits involved in the gate, and the idle spectator qutrit. To cancel the entangling phase between a spectator qutrit and a gate qutrit, we use an additional microwave drive on the idle spectator qutrit charge line. Additionally, we characterize and correct the local rotation on the idle qutrit using virtual-Z gates in both the $\{\ket{0},\ket{1} \}$ and $\{\ket{1},\ket{2} \}$ subspaces of the qutrit. Nonetheless, there often remains at least one subspace of the qutrit with a significant amount of spurious entangling phase. To visualize and characterize the coherent error in terms of entangling phase between a gate-qutrit and spectator qutrit, we perform the experiment shown in Figs.~\ref{fig:S1} a-c.

\section{Variation Distance in RCS vs Number of Identity Insertions}\label{sec:id_inser}

\begin{figure}
    \centering
    \includegraphics[width = 0.9\columnwidth]{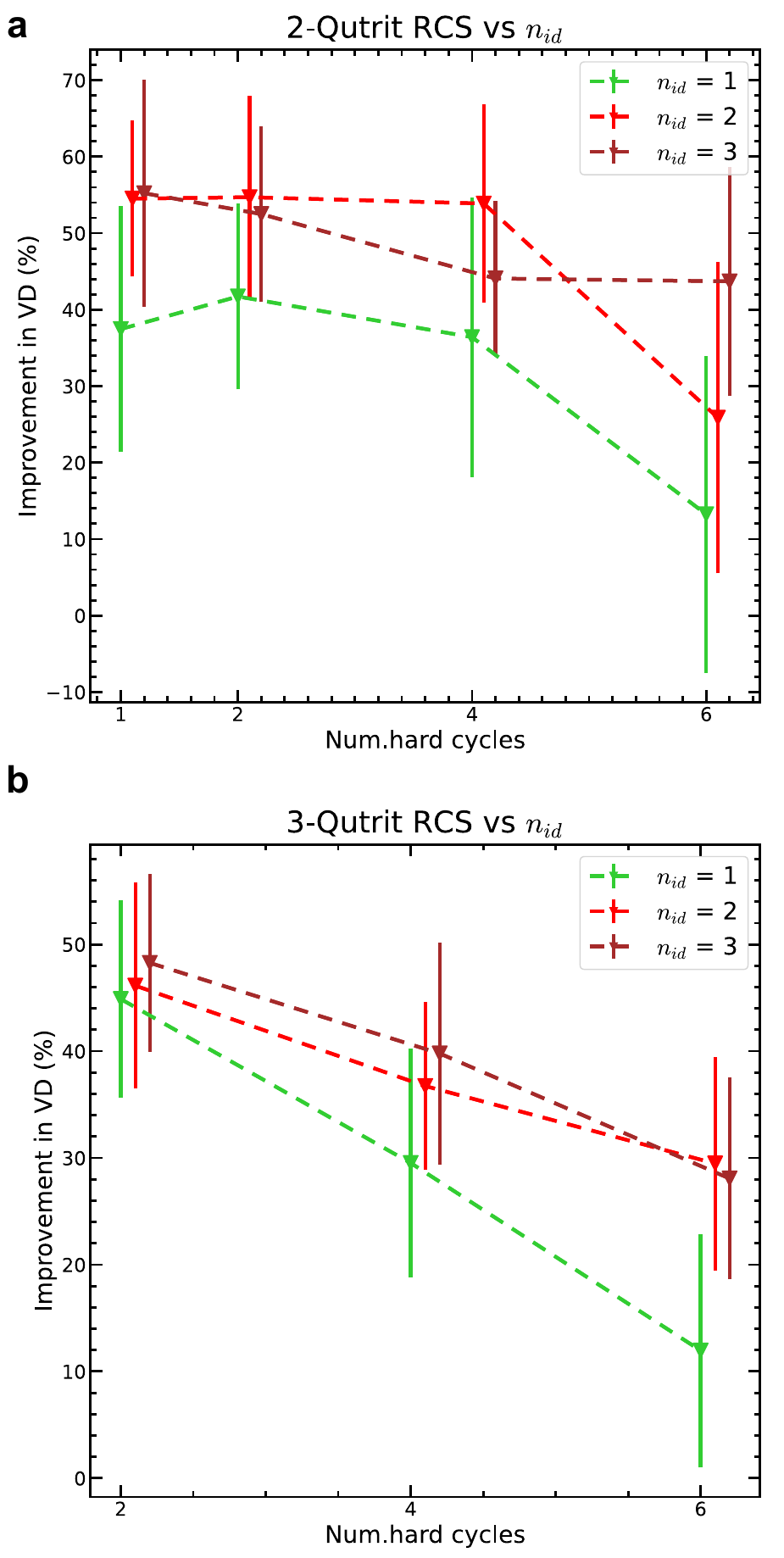}
    \caption{Qutrit RCS Imporvements with Variable Insertions of the Identity. (a) 2-qutrit RCS data showing improvements in the variation distance as a function of the number of identity insertions ($n_{id}$) as a function of circuit depth. (b) 3-qutrit RCS data.}
    \label{fig:S2}
\end{figure}

To effectively mitigate stochastic noise in circuits that are operated at deeper depths, we expect that more insertions of the identity will be required. This is a consequence of the fact that in deeper circuits, there is more decoherence, effectively raising the noise floor. To extrapolate beyond this noise floor, it is necessary to amplify additional noise in the form of more identity insertions. To confirm this behavior and find the ideal number of identity insertions ($n_{id}$), we study the problem of qutrit RCS outlined in the main text, and compare our results for $n_{id} = \{1,2,3 \}$ with 20 RC randomizations in each case. The results for these experiments are found in Fig.~\ref{fig:S2}. 

Consistent with our intuition, we find no statistically significant difference between different choices of $n_{id}$ for low depth circuits. However, at longer circuit depths, we observe significant improvements for $n_{id} = 2, 3$ over $n_{id} = 1$. This motivates the choice to use $n_{id} = 3$ for both 2 and 3-qutrit RCS data presented in the main text, and also motivates our decision to use $n_{id} = 1$ for the 3-qutrit GHZ experiment, as it only required two hard cycles.

\begin{figure*}
    \centering
    \includegraphics[width = 0.9\textwidth]{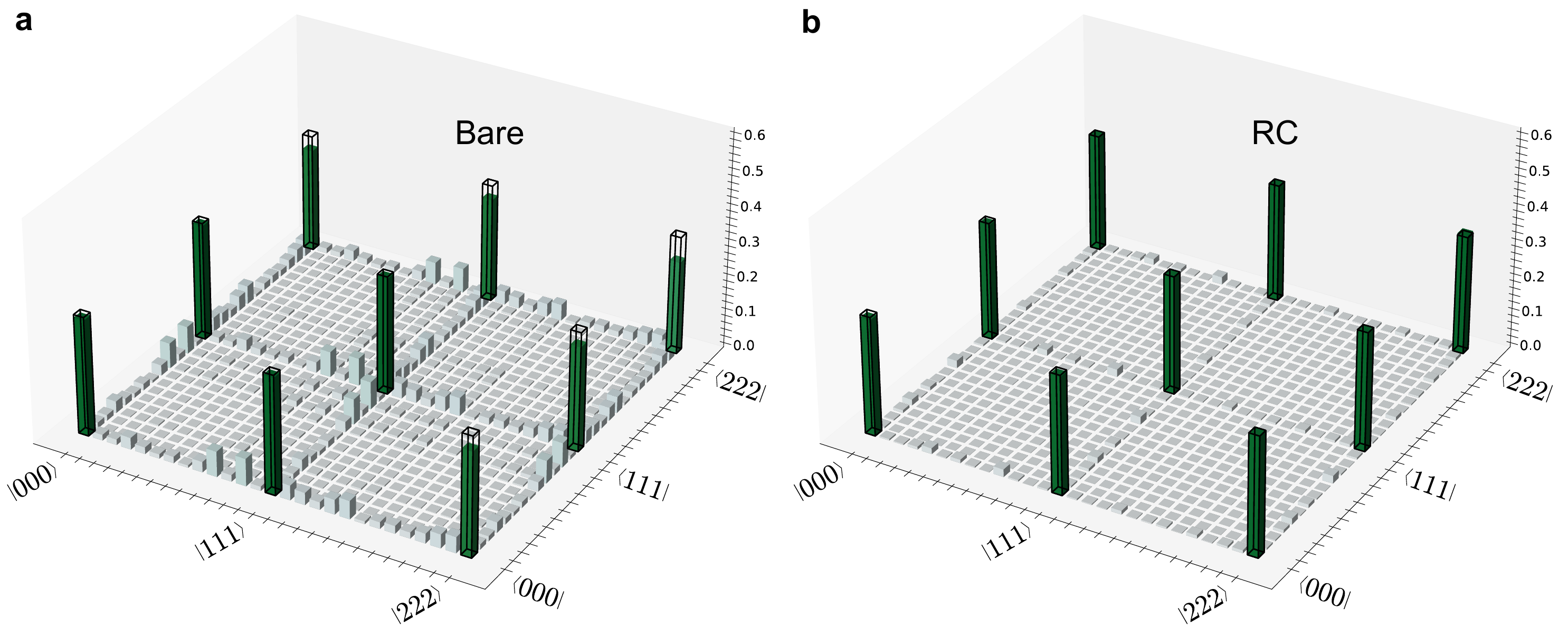}
    \caption{Purified GHZ Density Matrices with and without randomized compiling. (a) The purified density matrix in the bare case, with a state fidelity of $\mathcal{F} = 0.912$. (b) The purified density matrix in the 20 randomizations RC case, with $\mathcal{F} = 0.998$.} 
    \label{fig:S3}
\end{figure*}

\section{Calculating the Proportion of Coherent Errors in Qudit PTMs}\label{sec:coh_fraction}

In Fig.~\ref{fig:fig2}d, we show the numerical results of twirling away (using qudit RC) off-diagonal elements (or coherent errors) in qudit PTMs. Here, we briefly comment on how one calculates the proportion of coherent errors present in a qudit PTM, following Ref.~\cite{Carignan_Dugas_2019}.

Consider a quantum error channel with PTM denoted as $\mathcal{E}$. The
process fidelity is defined as the normalized trace of the ($d^2 \times d^2$) 
PTM:
\begin{align}
    \fidP{\mathcal{E}} := \trace \mathcal{\mathcal{E}}/d^2
\end{align}
Let's define the decoherent process fidelity of $\mathcal{E}$ 
as
\begin{align}\label{def:fiddecoh}
    \fiddecoh{\mathcal{E}} := \sqrt{\trace \mathcal{\mathcal{E}^\dagger\mathcal{E}}/d^2} = \|\mathcal{E}\|_F/d~.
\end{align}
In Ref.~\cite{Carignan_Dugas_2019}, it is shown that as long as 
the error channel is reasonably close to the identity, that is 
if $\fidP{\mathcal{E}}>1/2$ and $\fiddecoh{\mathcal{E}} > 1/\sqrt{2}$, the 
error channel has a well-defined coherent-decoherent polar decomposition:
\begin{align}\label{eq:polar}
    \mathcal{E} = \mathcal{U} \mathcal{D}~,
\end{align}
where $\mathcal{U}$ is a coherent (i.e.~unitary) error process and $\mathcal{D}$ is a purely
decoherent process. The precise definition of a decoherent channel is 
elaborated and justified in Ref.~\cite{Carignan_Dugas_2019}. 

In realistic physical scenarios, the process fidelity can be expressed as a 
simple product:
\begin{align}\label{eq:fid_prod}
    \fidP{\mathcal{E}} = \fidP{\mathcal{U}} \fidP{ \mathcal{D}} + h.o.~,
\end{align}
where the higher order term ($h.o.$) is of second order in the infidelity, 
$O\left((1-\fidP{\mathcal{E}})^2\right)$. For reference, \cref{eq:fid_prod} 
does not hold in pathological cases where a significant part of the error 
process originates from specially crafted high-body interactions (e.g.~
Hamiltonian/Lindbladian terms that are made of tensor products of a large 
number of subsystems). A discussion of these pathological cases is 
included in Ref.~\cite{Carignan_Dugas_2019}.

Notice that by combining \cref{def:fiddecoh} and \cref{eq:polar}, we get:
\begin{align}\label{eq:decoh_fid_reexpress}
    \fiddecoh{\mathcal{E}}  & = \fiddecoh{\mathcal{U} \mathcal{D}} \notag \\
    & = \|\mathcal{U} \mathcal{D}\|_F/d \notag \\
    & = \| \mathcal{D}\|_F/d \notag \\
    & = \fiddecoh{\mathcal{D}} \notag \\
    & = \fidP{\mathcal{D}} + O\left((1-\fidP{\mathcal{D}})^2\right)
\end{align}
where the last line comes from the fact that non-pathological 
decoherent errors obey (\cite{Carignan_Dugas_2019})
\begin{align}
    \trace \mathcal{D}^\dagger \mathcal{D} / d^2  = \left( \trace \mathcal{D} /d^2 \right)^2 +O\left((1-\fidP{\mathcal{D}})^2\right)~.
\end{align}
The above is essentially a corollary from the fact that decoherent errors build up according to 
a multiplicative decay. 

By substituting \cref{eq:decoh_fid_reexpress} in \cref{eq:fid_prod}, we get, up to second order in the infidelity,
\begin{align}\label{eq:fid_prod_update}
    \fidP{\mathcal{E}} = \fidP{\mathcal{U}} \fiddecoh{\mathcal{E}},
\end{align}
or in terms of infidelity (up to second order in the infidelity)
\begin{align}\label{eq:infid}
    \underbrace{1-\fidP{\mathcal{E}}}_{\rm Tot.~infid.} = \underbrace{1-\fidP{\mathcal{U}}}_{\rm Coh.~infid.} + \underbrace{1- \fiddecoh{\mathcal{E}}}_{\rm Decoh.~infid.}~.
\end{align}

As such, the contribution of coherence to the process infidelity is given by
(up to the second order in the infidelity):
\begin{align}\label{eq:cohinfid}
    1-\fidP{\mathcal{U}} = \fiddecoh{\mathcal{E}}-\fidP{\mathcal{E}}~.
\end{align}
The relative coherent contribution to the infidelity is therefore obtained via
\begin{align}
     \frac{\fiddecoh{\mathcal{E}}-\fidP{\mathcal{E}}}{1- \fidP{\mathcal{E}}} = \frac{\trace{\mathcal{E}}/d^2-d\|\mathcal{E}\|_F/d}{1- \trace{\mathcal{E}}/d^2}~.
\end{align}
The above is the formula used in Fig.~\ref{fig:fig2}d to quantify the effect of twirling on the error channel.

\section{Purification of GHZ Density Matrices}\label{section:purification}

In the main text, it was noted that our experiments studying multipartite qutrit entanglement yielded experimental evidence of RC tailoring coherent errors into stochastic noise. Here, we flesh out this point further. To purify the experimentally reconstructed GHZ density matrices, we use the Mcweeny purificaton formula,
\begin{equation}
    \rho_{n+1} = 3\rho_{n}^2 -2\rho_{n}^3 .
\end{equation}
We repeat the purification for $n=20$ steps and find that this converges to the nearest idempotent pure state. In Fig.~\ref{fig:S3}, we plot the purified bare density matrix, as well as the purified density matrix measured using 20 randomizations under RC. Notably, the state fidelity of the purified density matrix improved from 0.912 in the bare case, to 0.998 when employing RC. This is a strong indication that initially our system had significant coherent errors, in which case the main contribution to the infidelity was due to angle errors --- as opposed decoherence errors --- in which case state purification would not recover the ideal state, but rather some other pure state that is still offset from the ideal state. However, when employing RC, these angle errors are tailored to decoherence errors, leading to a reduction in the length of the generalized Bloch vector, which can be corrected via numerical state purification. Importantly, as full state tomography has an exponential overhead with large system sizes, we do not consider purification to be a feasible tool going forward, but we do note its value here at demonstrating the power of RC for noise tailoring in qudit systems.
\bibliographystyle{ieeetr}
\bibliography{bibliography}

\end{document}